\documentclass[sigconf]{acmart}
\AtBeginDocument{%
  }

\setcopyright{acmlicensed}
\copyrightyear{2026}
\acmYear{2026}
\setcopyright{cc}
\setcctype{by}
\acmConference[IVA 2026]{ACM International Conference on Intelligent Virtual Agents}{September 07--11, 2026}{Puebla, Mexico}
\acmBooktitle{ACM International Conference on Intelligent Virtual Agents (IVA 2026), September 07--11, 2026, Puebla, Mexico}
\acmDOI{10.1145/3806774.3832791}
\acmISBN{979-8-4007-2647-7/2026/09}

\begin{document}

\title{Evaluating Prompt Robustness in Text-to-Audio Systems for Adaptive Virtual Agents and Game Soundtracks}

\author{Jiahui Wu}
\email{wuj17@rpi.edu}
\orcid{0009-0009-4839-5957}
\correspondingauthor
\affiliation{%
  \institution{Rensselaer Polytechnic Institute}
  \city{Troy}
  \state{New York}
  \country{USA}
}

\author{Mei Si}
\email{sim@rpi.edu}
\orcid{0000-0001-8642-8806}
\correspondingauthor
\affiliation{%
  \institution{Rensselaer Polytechnic Institute}
  \city{Troy}
  \state{New York}
  \country{USA}
}

\begin{abstract}
Recent text-to-audio models enable adaptive game soundtracks, but small prompt changes can cause abrupt musical shifts. We evaluate MusicGen-small, MusicGen-large, and Stable Audio 2.5 under Minimal Lexical Substitution, Intensity Shifts, and Structural Rephrasing using log-Mel distance, MFCC/chroma-DTW, and CLAP similarity. Stable Audio 2.5 achieves the lowest pooled acoustic distances and the highest audio–audio CLAP similarity under structural rephrasing, while MusicGen-large has the highest audio–audio CLAP similarity under lexical substitutions and intensity shifts. Stable Audio 2.5 also shows the greatest between-seed variation in prompt-to-audio alignment, demonstrating the importance of multi-seed robustness evaluation for adaptive game audio.

\end{abstract}

\ccsdesc[500]{Applied computing~Sound and music computing}
\ccsdesc[300]{Applied computing~Computer games}
\ccsdesc[300]{Human-centered computing~Auditory feedback}
\ccsdesc[300]{General and reference~Evaluation}

\keywords{Adaptive Game Audio, Text-to-Audio Generation, Generative Audio Models, Prompt Robustness, Semantic Perturbation Analysis, Audio Similarity Evaluation}

\maketitle

\section{Introduction}

Intelligent virtual agents increasingly adapt their verbal, nonverbal, and affective behavior to changing interaction contexts. In games and simulations, an agent may respond to its emotional state, player actions, narrative progression, or changes in the environment. Generative audio provides an additional channel through which an agent-driven system can communicate emotion, narrative significance, and gameplay state.

Text-to-audio generation is promising for this purpose because an agent architecture can translate symbolic or language-based state descriptions directly into music or sound. An affect model might describe an agent as ``anxious,'' while a narrative planner characterizes the same situation as ``tense infiltration.'' Such descriptions may be produced by different system components or generated automatically from the current interaction state. A text-to-audio model could then produce music suited to that state without requiring designers to author audio for every possible condition.

However, this integration creates a reliability problem. A stable state may be described using different words or sentence structures, while evolving states may be expressed at adjacent affective-intensity levels. If these small differences cause large shifts in instrumentation, rhythm, harmony, or spectral texture, the generated soundtrack may suggest a state transition that did not occur. This is not only an audio-quality issue; it is a failure of multimodal coordination between the agent's represented state and its externally perceived expression.

Adaptive behavior therefore requires both responsiveness and continuity. Audio should change when an agent's emotional appraisal or gameplay context changes meaningfully, but different descriptions of a stable state should remain musically coherent. Gradual increases in fear or urgency should produce appropriately graded changes, while rephrasing an unchanged event should not substantially alter the expressed mood. Because musical affect depends on interacting cues such as tempo, dynamics, mode, brightness, and instrumentation \cite{b4,b5}, minor prompt differences may be amplified into perceptually important changes.

This issue is especially relevant to modular social-agent architectures, where agent state, language realization, and audio generation may be handled by separate components, with natural-language prompts serving as an interface between them. Prompt robustness therefore affects interoperability, portability, and reuse, as well as the suitability of generative audio for real-time human--agent interaction.

We evaluate whether semantically related prompts preserve audio continuity under Minimal Lexical Substitution (MLS), Intensity Shifts (IS), and Structural Rephrasing (SR). Prior work has shown that prompt-based models can react strongly to small wording changes \cite{b1,b2,b3}, but we still know little about how such variation affects audio used as part of an agent's adaptive behavior.

Recent text-to-audio systems, including CLAP-based audio--language representation learning \cite{elizalde2023clap}, AudioLDM \cite{b7}, Tango \cite{b8}, and MusicGen \cite{b9}, have substantially improved text-conditioned audio generation. However, generation quality and prompt adherence alone do not show whether these systems behave consistently across equivalent or gradually changing descriptions of an agent or gameplay state. 

We evaluate three models with different scales and generation approaches: MusicGen-small, a 300M-parameter autoregressive Transformer that generates music from discrete EnCodec audio tokens \cite{b9,b10}; MusicGen-large, which uses the same architecture with approximately 3.3B parameters \cite{b11}; and Stable Audio 2.5, a system for music and sound-effect generation \cite{b12}. We compare the outputs at three complementary levels. Log-Mel distances capture changes in spectral energy and texture, MFCC- and chroma-based DTW measure timbral and harmonic similarity while accounting for timing differences, and CLAP embeddings capture higher-level semantic and affective similarity \cite{davis1980mfcc,muller2007dtw,fujishima1999chroma,elizalde2023clap}. No single metric fully represents perceived musical continuity, so we use them together to identify whether prompt variation affects acoustic realization, musical structure, or high-level meaning.

\section{Models and Experimental Materials}

We construct a controlled prompt dataset with three types of variation: Minimal Lexical Substitution (MLS), Intensity Shifts (IS), and Structural Rephrasing (SR). MLS replaces one affective descriptor with a near-synonym, IS organizes descriptors into four ordered intensity levels, and SR expresses the same game event using a different sentence structure. MLS and IS use the template ``Generate music that feels [descriptor],'' while SR uses complete descriptions of game-like events.

The dataset contains 75 prompt groups: 30 MLS pairs, 15 IS groups with four ordered descriptors each, and 30 SR pairs. For IS, we compare adjacent intensity levels, producing 45 comparisons. In total, the dataset includes 30 MLS, 45 IS, and 30 SR comparisons. To keep outputs comparable across models and conditions, for each model, every prompt is generated using six random seeds (0, 1, 2, 3, 4, and 42), producing six approximately 5-second clips per prompt. For each seed, the random-number generators were reset immediately before each single-prompt generation, ensuring that matched prompt variants began from the same sampling state. Each condition contains 60 prompts, yielding 360 clips per model–condition and 3,240 clips overall. Table~\ref{tab:game_examples} shows representative examples.

\begin{table}[t]
\centering
\caption{Representative prompt variations across game-audio scenarios. Different scenarios are shown to illustrate dataset coverage; each variation type is applied across multiple scenarios in the full dataset.}
\label{tab:game_examples}
\begin{tabular}{p{0.16\linewidth} p{0.21\linewidth} p{0.5\linewidth}}
\hline
\textbf{Condition} & \textbf{Adaptive Audio Scenario} & \textbf{Controlled Prompt Variants} \\
\hline

MLS & Stealth / suspense &
tense $\leftrightarrow$ anxious \\

MLS & Horror / danger &
nervous $\leftrightarrow$ uneasy \\

MLS & Action / energy &
energetic $\leftrightarrow$ lively \\

\hline

IS & Threat escalation &
Slightly worried $\rightarrow$ Concerned $\rightarrow$ Afraid $\rightarrow$ Terrified \\

IS & Safe-zone transition &
Calm $\rightarrow$ Relaxed $\rightarrow$ Composed $\rightarrow$ Serene \\

IS & Combat pressure &
Restless $\rightarrow$ Agitated $\rightarrow$ Frustrated $\rightarrow$ Overwhelmed \\

\hline

SR & Combat victory &
The hero wins the battle. \newline
The battle is won by the hero. \\

SR & Hidden-path discovery &
They discover the hidden path. \newline
The hidden path is discovered by them. \\

SR & Military movement &
The soldiers march forward. \newline
Forward march the soldiers. \\

\hline
\end{tabular}
\end{table}

\section{Evaluation}

We evaluate robustness at three complementary levels: spectral similarity, temporal–musical structure, and semantic similarity.

First, we compute L1 and root-mean-square error (RMSE) between duration-matched log-Mel spectrograms. Mel features summarize spectral energy over perceptually motivated frequency bands and are widely used in audio analysis \cite{davis1980mfcc}. Lower distances indicate greater similarity in spectral texture, brightness, and energy distribution.

Second, we compute Dynamic Time Warping (DTW) costs over MFCC and chroma sequences. MFCCs represent spectral-envelope and timbral characteristics \cite{davis1980mfcc}, while chroma features summarize pitch-class and harmonic content \cite{fujishima1999chroma}. DTW aligns sequences before comparison, allowing for small timing or tempo differences \cite{muller2007dtw}. Lower costs indicate greater timbral and harmonic consistency after alignment.

Third, we measure high-level similarity using CLAP audio embeddings. CLAP maps audio and text into a shared representation space learned from audio--language pairs \cite{elizalde2023clap}. We compute cosine similarity between normalized clip embeddings. Higher cosine similarity indicates greater similarity in the learned semantic space, including scene, mood, and affect.

These metrics capture different forms of fragility: two clips may preserve the same broad mood while differing in texture or harmony, or remain acoustically similar while conveying different meanings. We therefore interpret them together. Because log-Mel and DTW distances have no universal thresholds, we compare them across models and prompt conditions within the same pipeline.

Across six seeds, Figure 1 pools matched-seed audio–audio comparisons, yielding 180 comparisons for MLS and SR and 270 for IS per model. For Figure 2, we compute prompt-to-audio CLAP cosine similarity for each clip and average the 60 scores within each model–condition–seed combination, producing six seed-level means per model and condition. Box plots summarize these means: the center line indicates the median, the box indicates the interquartile range, and the whiskers indicate the minimum and maximum.

For each MusicGen model–condition–seed combination, audio–audio CLAP scores were averaged across prompt pairs. MusicGen-small and MusicGen-large were compared separately under MLS, IS, and SR using two-tailed paired-samples t-tests on seed-level means matched by seed identifier (n=6 pairs, df=5, $\alpha$=.05). $SE_{diff}$ denotes the standard error of the mean paired difference. Levene’s tests were additionally reported as supplementary examinations of differences in the models’ marginal seed-level variances. All reported p-values are two-tailed and unadjusted for the three condition-specific comparisons.


\section{Results}

\subsection{Audio Similarity Results}

Figure 1 summarizes the audio–audio similarity results pooled across six random seeds. Stable Audio 2.5 produces the lowest pooled L1, RMSE, MFCC-DTW, and chroma-DTW values under all three prompt conditions. For audio–audio CLAP similarity, MusicGen-large achieves the highest values under MLS and IS, whereas Stable Audio 2.5 is highest under SR. MusicGen-large exceeds MusicGen-small under MLS and IS, but falls below MusicGen-small under SR. These results reinforce that semantic consistency and acoustic continuity capture complementary aspects of robustness.

Two-tailed paired-samples t-tests compared MusicGen-large and MusicGen-small across six matched seeds. Under MLS, Levene’s test did not indicate significantly unequal marginal variances, $F(1,10)=3.49, p>.05$. MusicGen-large ($M=.79, SD=.02$) achieved significantly higher audio–audio CLAP similarity than MusicGen-small ($M=.64, SD=.04$), $SE_{diff}=.02, t(5)=7.97, p<.01$. Under IS, Levene’s test was also nonsignificant, $F(1,10)=0.13, p=.73$, and MusicGen-large ($M=.81, SD=.03$) significantly outperformed MusicGen-small ($M=.65, SD=.04$), $SE_{diff}=.025, t(5)=6.42, p<.01$. Under SR, Levene’s test was nonsignificant, $F(1,10)=0.09, p=.77$, and the difference between MusicGen-large ($M=.61, SD=.02$) and MusicGen-small ($M=.64, SD=.03$) was not significant, $SE_
{diff}=.02, t(5)=-2.01, p>.05$. These results suggest that increased model capacity improves semantic stability for lexical substitutions and graded intensity changes.

\begin{figure}[htbp]
\centerline{\includegraphics[width=1.0\linewidth]{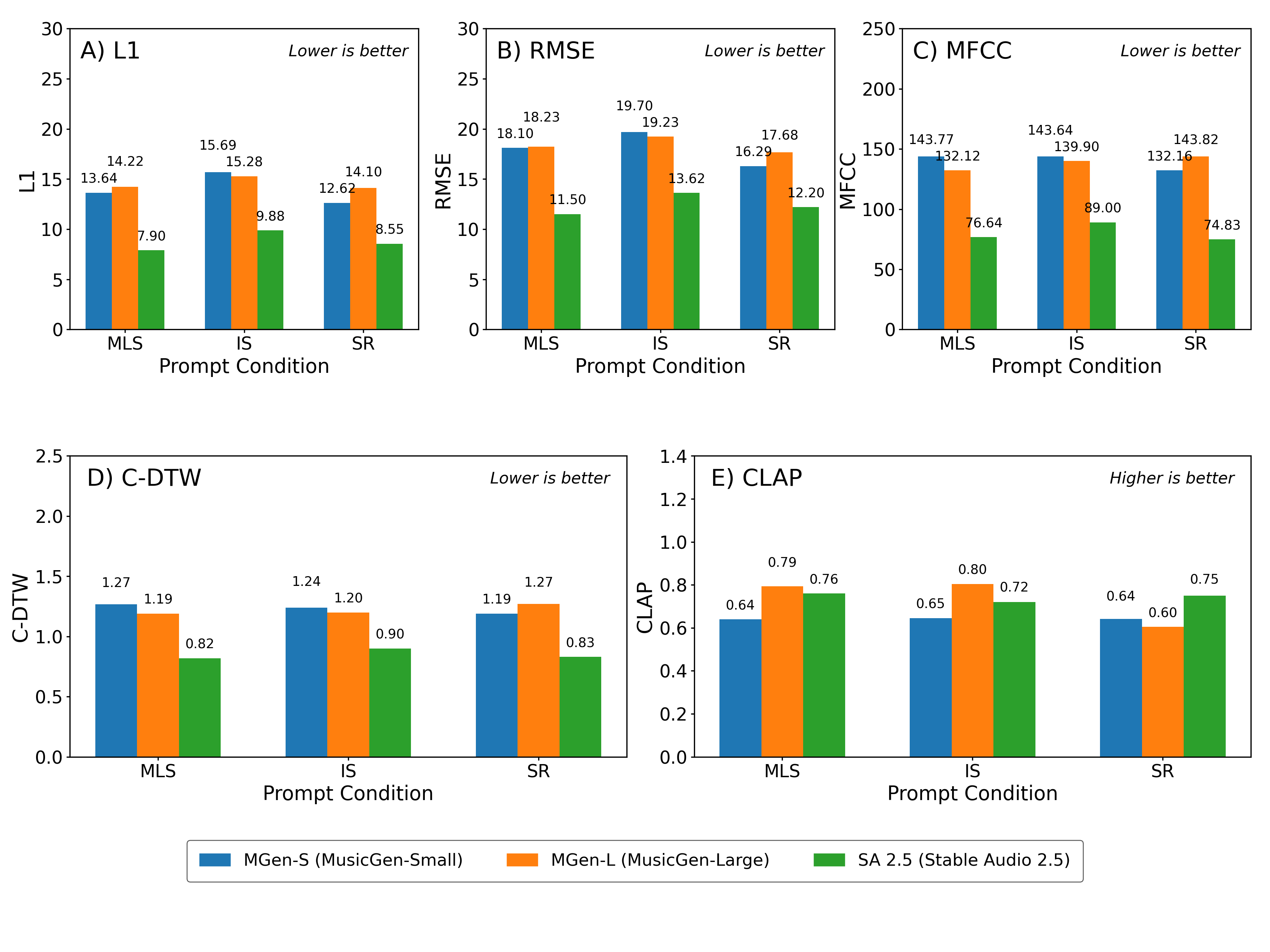}}
\caption{Audio–audio similarity results across models and prompt variations}
\Description{Audio similarity results for MusicGen-small, MusicGen-large, and Stable Audio 2.5 under Minimal Lexical Substitution, Intensity Shifts, and Structural Rephrasing.}
\label{fig:audio_similarity}
\end{figure}

\subsection{Prompt-to-Audio Alignment}

Figure 2 summarizes the distributions of seed-level mean prompt-to-audio CLAP similarity. Stable Audio 2.5 has the highest median under MLS, IS, and SR, while every model achieves its highest median under SR. Stable Audio 2.5 also exhibits the greatest between-seed variation, particularly under IS, whereas MusicGen-large shows the narrowest seed-level distribution across all three conditions. Thus, Stable Audio 2.5's stronger prompt adherence is accompanied by greater sensitivity to the random seed.

\begin{figure}[htbp]
\centerline{\includegraphics[width=1.0\linewidth]{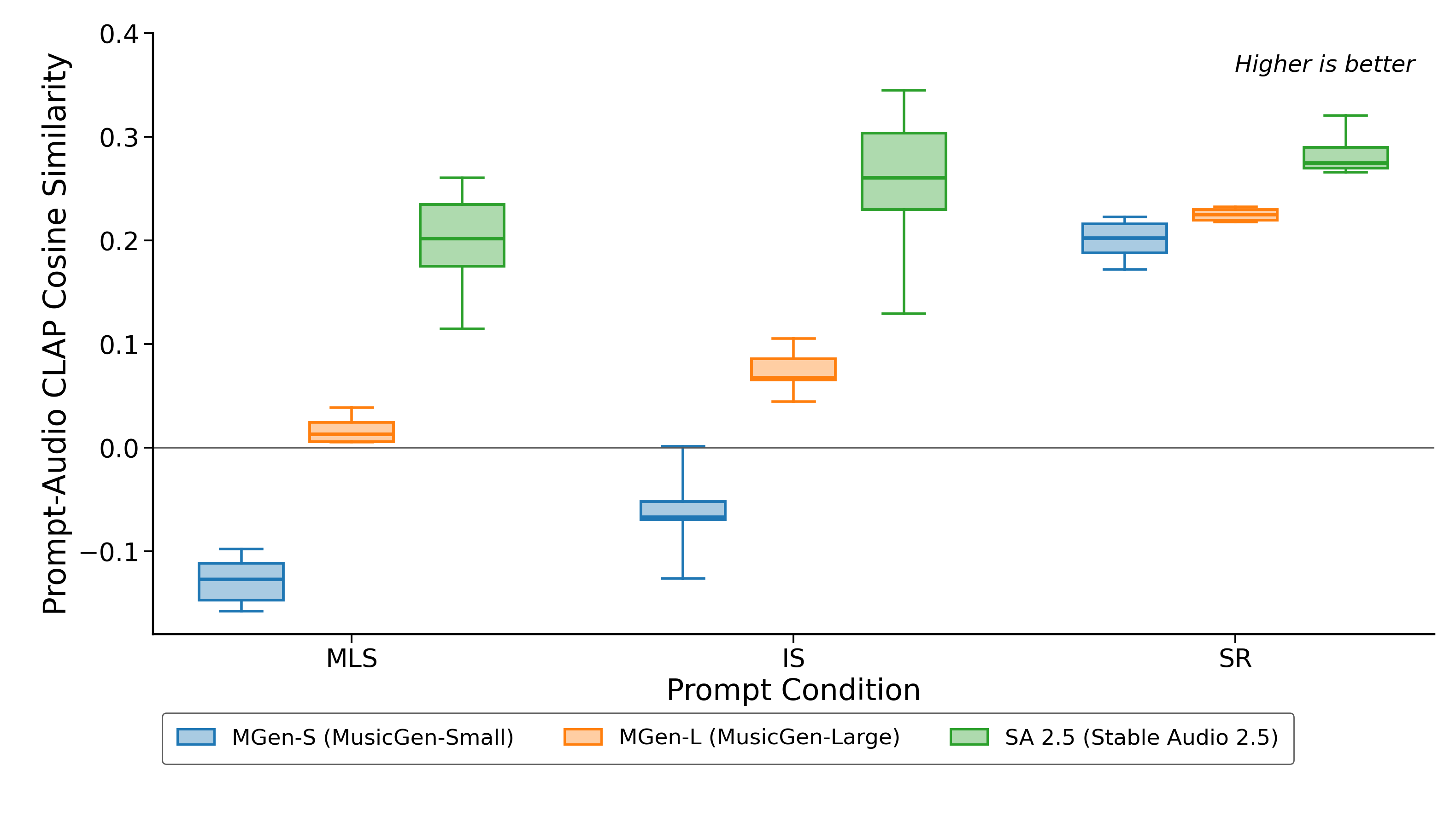}}
\caption{Seed-level prompt-to-audio CLAP cosine-similarity distributions across six random seeds. Boxes show the median and interquartile range; whiskers show the minimum and maximum.}
\Description{Cross-seed variation in prompt-to-audio CLAP cosine similarity. Each bar contains six seed-level means, with each seed mean computed from 60 clips. Points represent individual seeds, boxes show the interquartile range and median, and diamonds indicate cross-seed means.}
\label{fig:prompt_audio}
\end{figure}

\subsection{Qualitative Listening Comparisons}

Representative seed-0 pairs illustrate how the quantitative differences sound. Under IS, MusicGen-large's ``Agitated'' and ``Frustrated'' outputs retain a gloomy, restless character (CLAP similarity $=0.93$), with tonal changes reflecting the intended intensity shift. Under MLS, MusicGen-small's ``cheerful'' and ``delighted'' outputs convey happiness but differ in rhythmic density and frequency emphasis (log-Mel L1 $=15.52$). Under SR, Stable Audio 2.5's active and passive window descriptions remain quiet and slow (MFCC-DTW $=88.57$). Together, these examples show that prompt variants can preserve broad affect while changing their acoustic realization.

\section{Discussion, Limitations, and Future Work}

Our six-seed evaluation shows that no model is uniformly robust to small, meaning-preserving prompt variations. Stable Audio 2.5 yields the lowest acoustic distances and strongest median prompt alignment but the greatest between-seed variability. MusicGen-large improves audio–audio CLAP consistency over MusicGen-small under MLS and IS, but not SR, supporting multi-seed evaluation with complementary measures.

In adaptive game and virtual-agent systems, such instability may disrupt emotional continuity or imply nonexistent state changes. Pre-deployment testing can identify prompts causing large spectral, temporal, or semantic shifts. Potential remedies include prompt normalization, constrained templates, equivalent-prompt fine-tuning, and output reranking.

Limitations include the controlled prompt set, short clips, three models, author-defined intensity sequences, missing same-prompt/different-seed and unrelated-prompt baselines, and no formal listening study.

Future work should enforce consistency across equivalent and gradually changing prompts, use structured state controls, and apply uncertainty estimation or multi-sample reranking to improve stability while preserving meaningful musical variation.

\bibliographystyle{ACM-Reference-Format}
\bibliography{references}


\end{document}